\documentclass[preprint,aps,prb,endfloats]{revtex4}

\usepackage{epsfig}

\begin{document}

\title{Sodium atoms and clusters on graphite: a density functional study}

\author{K. Rytk\"onen$^1$, J. Akola$^2$, and M. Manninen$^1$}
\affiliation{$^1$Department of Physics, University of
Jyv\"askyl\"a, FIN-40351 Jyv\"askyl\"a, Finland}
\affiliation{$^2$Institut f\"ur Festk\"orperforschung,
Forschungszentrum J\"ulich, D-52425 J\"ulich, Germany}

\date{\today}

\begin{abstract}
Sodium atoms and clusters ($N\le5$) on graphite (0001) are studied
using density functional theory, pseudopotentials and periodic
boundary conditions. A single Na atom is observed to bind at a
hollow site 2.45 {\AA} above the surface with an adsorption energy
of 0.51 eV. The small diffusion barrier of 0.06 eV indicates a
flat potential energy surface. Increased Na coverage results in a
weak adsorbate-substrate interaction, which is evident in the
larger separation from the surface in the cases of Na$_3$, Na$_4$,
Na$_5$, and the (2$\times$2) Na overlayer. The binding is weak for
Na$_2$, which has a full valence electron shell. The presence of
substrate modifies the structures of Na$_3$, Na$_4$, and Na$_5$
significantly, and both Na$_4$ and Na$_5$ are distorted from
planarity. The calculated formation energies suggest that
clustering of atoms is energetically favorable, and that the open
shell clusters (e.g. Na$_3$ and Na$_5$) can be more abundant on
graphite than in the gas phase. Analysis of the lateral charge 
density distributions of Na and Na$_3$ shows a charge transfer of
${\sim0.5}$ electrons in both cases. 

\end{abstract}

\maketitle


\section{Introduction}
Graphite is a semimetal that is widely utilized in experimental
surface physics. The planar geometry and weak van der Waals -type
interlayer interaction make it possible to split flat, chemically
inert, and clean graphite (0001) surfaces (highly oriented
pyrolytic graphite, HOPG), which are ideal for studying adsorption
layers and clusters. The electronic 2D semimetal properties of
graphite are well-known both experimentally and theoretically. For
instance, density functional theory (DFT) has provided
information about the valence charge density, electronic density
of states, band structure, elastic constants, and equation of
state. \cite{Cha91,Cha92,Boe97,Kga03}

An interesting research field considers alkali metal atoms and
clusters on graphite. Reactivity and metallic properties make
alkali metals exciting both for nanotechnological applications and
basic science, and the properties of adsorbed alkali metal atoms on
HOPG evolve as a function of coverage. Initially, a dispersed and
highly polarized phase (``correlated liquid'') is found where
alkali atoms maintain a maximum distance between each other. After
a critical density of adatoms is reached, a nucleation to more
closely packed configurations (islands) occurs. \cite{Hun98}
Alkali metals seem to have a higher charge transfer to HOPG with
lower coverage, and an increase in adatom density tends to
re-organize the charge into the alkali metal layer forming a
two-dimensional metallic state that has a small surface corrugation
and is almost decoupled from the substrate.
\cite{Hun98,Whi94,Anc93,Lam98,Bre02} It has been proposed that
alkali-metal-plated graphite could have practical applications as
a substrate in studying normal and superfluid He films.
\cite{Whi94,Che91}

Despite the similar electronic structure of alkali metals,
deviations in island formation and interaction with HOPG are
observed as the atomic number increases. While lithium
atoms either intercalate between the graphene layers \cite{Kga03}
or form a planar incommensurate hcp superstructure on HOPG,
\cite{Hu84} it has been suggested that sodium nucleates only in
buckled (110) bcc overlayers. \cite{Bre01,Bre03} The larger alkali
atoms (K,Ru,Cs) are found to intercalate via surface defects
\cite{Whi94,Bre02} or to adsorb in a ($2\times2$) phase occupying
hollow sites of the hexagonal substrate.
\cite{Whi94,Bre02,Hu86,Hun96} In addition, cesium can exist
in an incommensurate hexagonal or a more sparse
($\sqrt7\times\sqrt7$)R19.11$^\circ$ phase,
\cite{Whi94,Hu86,Hun96} and a dense
$(\sqrt{3}\times\sqrt{3})$R30$^{\circ}$ structure has been
proposed for potassium. \cite{Wu82} Obviously, the above
observations are related to the atomic radius and ionization
potential of the alkali atom in question, which affect both the
adatom-adatom and adatom-surface interactions.

The first experimental studies considered metal islands and
metal-layers on graphite. A more controlled treatment of
adsorbates is challenging, and it is difficult to study separated
atoms and size-selected small clusters. Contemporary experimental
techniques are able to deal with the practical difficulties such
as the substrate temperature, surface defects, kinetic energy of
cluster deposition (``soft-landing''), and cluster aggregation.
\cite{Pal03} Theoretical studies concern mostly single atoms
\cite{Lou00,Duf98a,Duf98b,Gal03} or atomic layers on graphite
formed by periodic boundary conditions. \cite{Anc93,Hjo98,Lam98}
However, research on supported clusters is needed because they
form a bridge between isolated atoms and ordered nanolayers, and
they may have nanotechnological importance (quantum dots,
catalysis). Several attempts to model small clusters and molecules
on HOPG have been made, 
\cite{Duf98a,Duf98b,Lam98,Mou95,Hag00,Hag01,Sor01,Gia03,Fer02} 
but the large number of substrate atoms and the semimetallic 
nature of graphite ({\bf k}-points) make reliable calculations 
very demanding.

Various theoretical methods are capable of studying metal atoms
and clusters on graphite (0001). In addition to deciding which
theoretical tools to use, a crucial question is how to model a
graphite surface, i.e., how many graphene layers are needed, how
large should the substrate be, and does the adsorbate change the
surface geometry? One approach is to place the metal cluster under
study onto an isolated hydrogen-terminated piece of graphite
(``cluster'') that mimics a continuous surface.
\cite{Lou00,Duf98a,Duf98b,Fer02} The question then is how large should
the graphite cluster be in order to get realistic results? On the
other hand, there is a problem in optimizing the geometry of the
substrate if several graphene layers are involved. This is due to
the fact that the layers are interlocked, and the system (Bernal
graphite, stacking ABAB) is not fully symmetric at the substrate
edges. With periodic boundary conditions one can describe, in
principle, a continuous infinite system (``slab'') in the lateral
dimensions. In this case, the problem is the distance between
adsorbate replicas, which should be large enough to exclude
charge density overlap. The large substrate that must be used
increases the computational cost greatly.

In the present work, a DFT method with periodic boundary
conditions has been used to model Na atoms and clusters ($N\le5$)
on HOPG. The substrate consisted of three graphene layers with 32
(60) carbon atoms each. It was found that the HOPG potential
energy surface (PES) is very flat with the hollow site of the
carbon hexagonal structure being preferred. Although alkali metal 
atoms tend to be more weakly bound to the surface when the 
coverage increases, this tendency is not so clear in small Na
clusters. The calculated cluster energetics favor clustering
processes on HOPG, and the stability of open shell clusters
(Na$_3$ and Na$_5$) is increased.

\section{Simulation methods}

The calculations have been performed using the Car-Parrinello molecular
dynamics (CPMD) package, \cite{CPMD} which is based on density
functional theory. The electron-ion interaction is described by
ionic pseudopotentials having the non-local, norm-conserving, and
separable form suggested by Troullier and Martins. \cite{TM91}
Periodic boundary conditions are employed, and the plane wave
basis has a kinetic energy cut-off of 70 Ry. The generalized
gradient-corrected approximation of Perdew, Burke and Ernzerhof
(PBE) \cite{PBE96} is adopted for the exchange-correlation energy
density. The electronic Hamiltonian is rediagonalized after each
geometry optimization step, and a finite temperature functional
($T=1000$ K) is used for the Kohn-Sham (KS) orbital occupancies
due to the small energy gap between the occupied and unoccupied
states (band gap). The ionic positions are optimized using a
conjugate gradient method until all the components of nuclear 
gradient are below 1{$\times$}{$10^{-4}$} a.u.

\begin{figure}
\epsfig{file=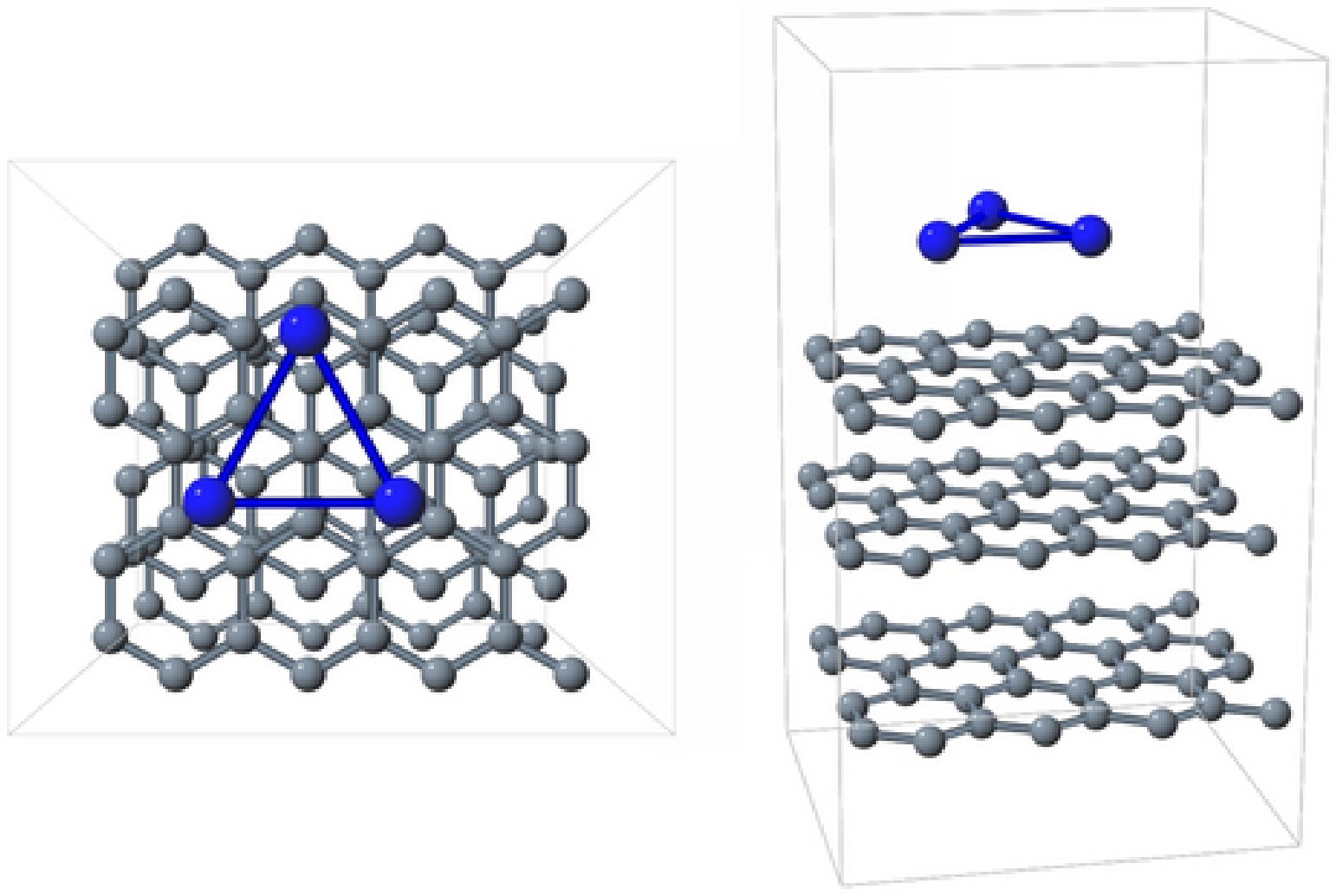}
\caption{Optimized Na$_3$-HOPG system shown from two perspectives.
The supershell size is 9.84$\times$8.53$\times$16.70 {\AA$^3$}. Each graphene
layer consists of 32 atoms.
}
\label{geom}
\end{figure}

We model two periodic substrates of Bernal graphite which consist
of three graphene layers (stacking ABA) in orthorhombic
supershells of 9.84$\times$8.53$\times$16.70 (96 C atoms) and
12.30$\times$12.79$\times$16.70 \AA$^3$ (180 C atoms). The smaller
substrate with Na$_3$ is shown in Fig. \ref{geom} from two
perspectives. Our tests for different numbers of graphene layers
have shown that at least three layers are needed in order to reach
a convergence in Na adsorption. \cite{Gr3} The spacing between the
layers is fixed to the experimental value 3.35 {\AA}, since the
PBE functional used has problems in describing weak van der
Waals-type interactions. \cite{Kga03} The choice of $z$-dimension
keeps the slab replica 10 {\AA} apart, which is sufficient for
most applications. However, a weak binding of Na$_2$ (and large
separation from the surface) forced us to use 2 {\AA} larger
spacing in this case. For Na$_4$ and Na$_5$ the interaction between
cluster replicas becomes significant in the smaller box, and a
larger substrate in $x$- and $y$-dimensions is needed, where the
minimum distance between the clusters is now 7.62 and 6.53 {\AA},
respectively.

Extensive tests for different numbers of {\bf k}-points have shown
that the simple $\Gamma$-point approximation is not reliable for
the systems studied. This is manifested by an artificial planar
elongation of the graphite hexagons during geometry optimization,
and is probably related to a strong downward dispersion of the
upper $\sigma$ bands at the $\Gamma$-point. \cite{Cha91,Cha92} The
problem does not occur with a 2$\times$2$\times$1 Monkhorst-Pack
{\bf k}-point mesh, and a variation of lateral dimension results
in a value 1.421 {\AA} for the C-C nearest neighbor distance
(experimental value 1.420 {\AA}). The Na adsorption energies
obtained show that a 5$\times$5$\times$1 mesh is adequate (see
also Table \ref{tab1}), whereas the forces are already well
converged for the 2$\times$2$\times$1 mesh. We have also tested
whether it is possible to use a smaller kinetic energy cutoff: In
comparison with 70 Ry, a calculation with 50 Ry yielded 0.17 eV
(33 \%) weaker binding of Na, and the related perpendicular
distance from the surface increased by 0.19 {\AA} (7.8 \%). This
shows that the computational cost cannot be reduced without losing
accuracy.

The effect of substrate relaxation has been studied by releasing the 
six nearest C atoms and reoptimizing the Na-HOPG system geometry. 
The changes are small (e.g. the C-C distance 1.424 {\AA}) which
validates the use of fixed substrate in real applications. A bench
mark calculation for Na-HOPG shows that the local spin density
(LSD) approximation does not improve the results because of the
large number of KS states involved and the nonmagnetic nature of
the system. The calculations below are done with spin-degenerate
KS orbitals except for isolated Na atom and Na clusters.

\section{Results}

\begin{figure}
\epsfig{file=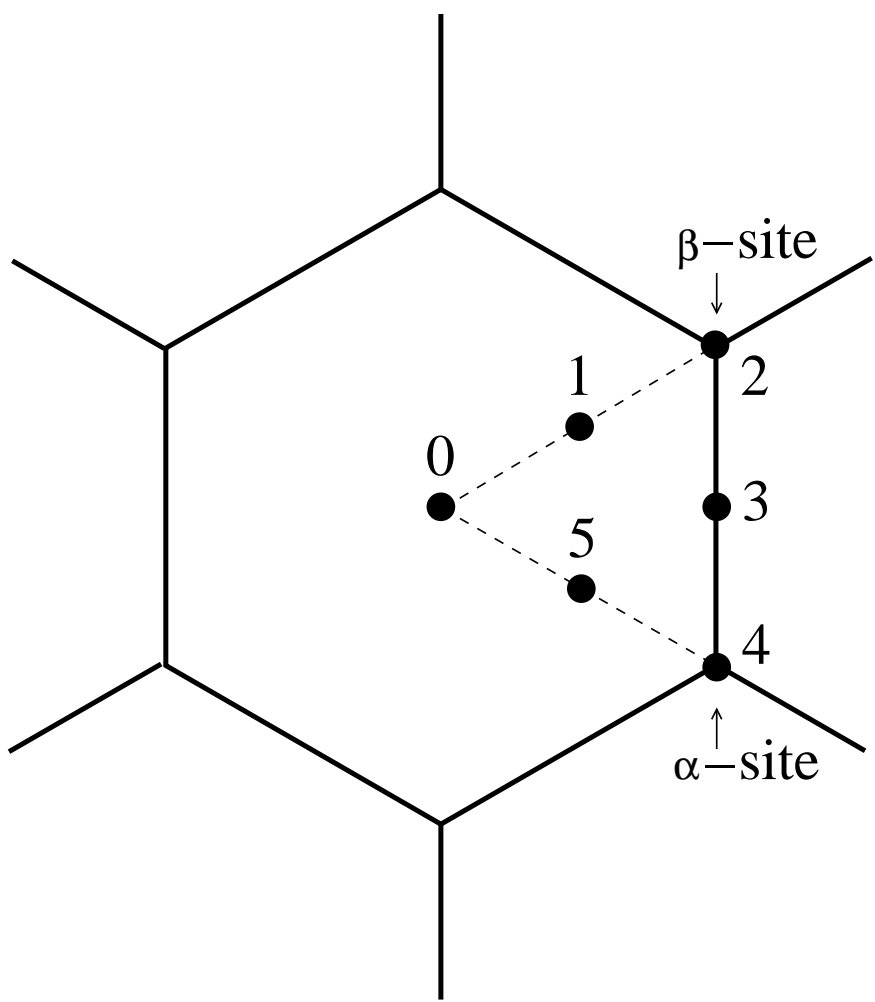}
\caption{Numbered locations of Na atom on top of a graphite hexagon.
}
\label{scheme}
\end{figure}

In order to map the potential energy surface of a Na-HOPG system,
we have optimized the Na atom position for different locations
along the surface (see Fig. \ref{scheme}). The results for
adsorption energy ($\Delta E_\bot$), separation from the surface
($d_\bot$), nearest carbon atom distances (Na-C), and carbon
coordination numbers ($N_C$) are presented in Table \ref{tab1}.
Here, we do not approach the real zero-density limit of Na, but
the atoms are distributed in 9.84$\times$8.53 {\AA} intervals due
to the periodic boundary conditions applied. \cite{Orth} Inclusion
of more {\bf k}-points enhances binding in a systematic way
yielding to an estimate of 0.51 eV for the energy minimum (point
0, 5$\times$5$\times$1 mesh). Comparison with other locations
shows only small deviations in $\Delta E_\bot$ and $d_\bot$,
indicating a flat potential energy surface with a maximum
variation of 0.07 eV. The points 2 and 4 above C$_\alpha$ and
C$_\beta$ (Fig. \ref{scheme}) give similar results, which causes
increased symmetry in the PES. These findings resemble the results
by Lamoen and Persson \cite{Lam98} who obtained
$\Delta E_\bot=0.52$ eV for a K-HOPG system and a small diffusion
barrier (variation 0.05 eV). No {\bf k}-points were used in these
calculations, but we expect a systematic shift in $\Delta E_\bot$ 
similar to the one we found.

Table \ref{tab2} shows the formation energetics of Na atoms and
clusters. The formation energy $\Delta E$ is divided into two
components: the binding energy of a free cluster or a separated
monolayer ($\Delta E_b$), and a term ($\Delta E_\bot$)
describing the adsorption energy. Three phases of Na-HOPG are
included in Table \ref{tab2}, where Na(I) refers to the initial
sparse system, Na(II) is a commensurate periodic structure with
twice as many Na atoms per unit shell as Na(I) (Na-Na separation
6.51 {\AA}), and Na(III) corresponds to the (2$\times$2) Na
monolayer with four times the coverage of Na(I) and hexagonal
symmetry. The phase Na(II) was encountered as a byproduct of
Na$_2$ stretching, and it corresponds to the maximal separation of
Na atoms allowed by the smaller supershell used. The effect of nearby 
Na atoms becomes clear in Na(III), where the separation from the
surface is 0.76 {\AA} greater. The loss in surface binding
is compensated by the interaction with other Na atoms, and the
resulting $\Delta E$ per atom is slightly larger than for Na(I). A
similar (2$\times$2) structure is stable for potassium,
\cite{Bre02} and theoretical studies have shown that K forms a
metallic state on HOPG. \cite{Whi94,Anc93,Lam98} Our results corroborate 
this finding, but -- in the case of Na monolayer -- the spacing between
Na atoms (4.92 {\AA}) does not match typical Na-Na distances (see
Na clusters in Table \ref{tab3}), and the energy difference with
the lower coverage phase Na(I) is relatively small.

The cluster formation energies in Table \ref{tab2} reveal
significant differences between individual clusters. Na$_2$ binds
only very weakly due to its closed valence electron shell, and the
dimer separation from the surface is 1 {\AA}  larger than for
Na$_3$ and Na$_4$. The same effect is apparent in the
$\Delta E_\bot$ values. It is interesting that the
deviation of $\Delta E$ for 2$\times$2$\times$1 and
5$\times$5$\times$1 {\bf k}-point meshes becomes smaller as the
distance between Na atoms and surface increases (see e.g. Na$_3$).
This implies changes in charge transfer. The larger substrate used
for Na$_4$ and Na$_5$ requires fewer {\bf k}-points to converge
the formation energy, which can be seen as nearly identical
$\Delta E$ values. The $\Delta E$ values of Na$_3$, Na$_4$, and
Na$_5$ are larger than for the (2$\times$2) Na monolayer, which
shows that the clustering of Na atoms is preferred.

\begin{figure}
\epsfig{file=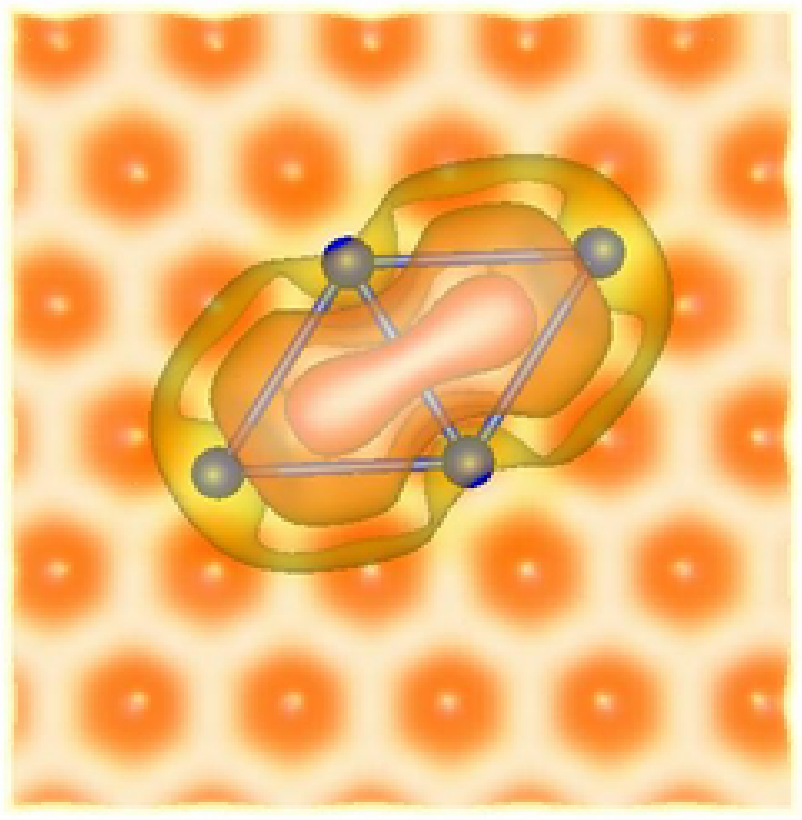}
\caption{Three density isosurfaces for Na$_4$-HOPG system. The corresponding
density values are 0.002 (yellow), 0.004 (orange), and 0.007 au (red),
respectively. The accumulated charges within the cluster are 1.77,
0.98, and 0.14 e, respectively.}
\label{vis_na4}
\end{figure}

The optimized cluster structures are related to the ground state
geometries of free Na clusters. For Na$_3$, Na$_4$, and Na$_5$ the
corresponding isomers are an isosceles triangle, a rhombus, and a
planar C$_{2v}$ isomer. The clusters are placed on HOPG in a way
that assumes that the hollow site (point 0) is energetically
favorable for each Na atom. The related bond distances, angles,
torsional angles, and distances from the surface are given in
Table \ref{tab3}. As mentioned above, Na$_2$ binds weakly, and this 
can also be seen in the very small change
in dimer bond length (0.02 {\AA}). For the other clusters changes
are more obvious: Na$_3$ adopts a geometry close to an equilateral
triangle with significant changes in bond lengths and angles.
Na$_4$ bends away from planarity (torsional angle 8.7$^\circ$),
and the bond lengths increase systematically, but the angles
remain close to the initial values. For Na$_3$ all atoms occupy
similar sites on top of point 5 (see Figs. \ref{geom},
\ref{scheme}), not far from hexagon centers. The atoms of Na$_4$
are coordinated with HOPG in two ways: the two corners of the
rhombus are above C atoms and bent towards the substrate (a result
contradictory to the PES of Na(I)), whereas the other two Na atoms
are close to the hexagon centers. The geometry and position of
Na$_4$ is shown in an electron density isosurface plot in Fig.
\ref{vis_na4}.

\begin{figure}
\epsfig{file=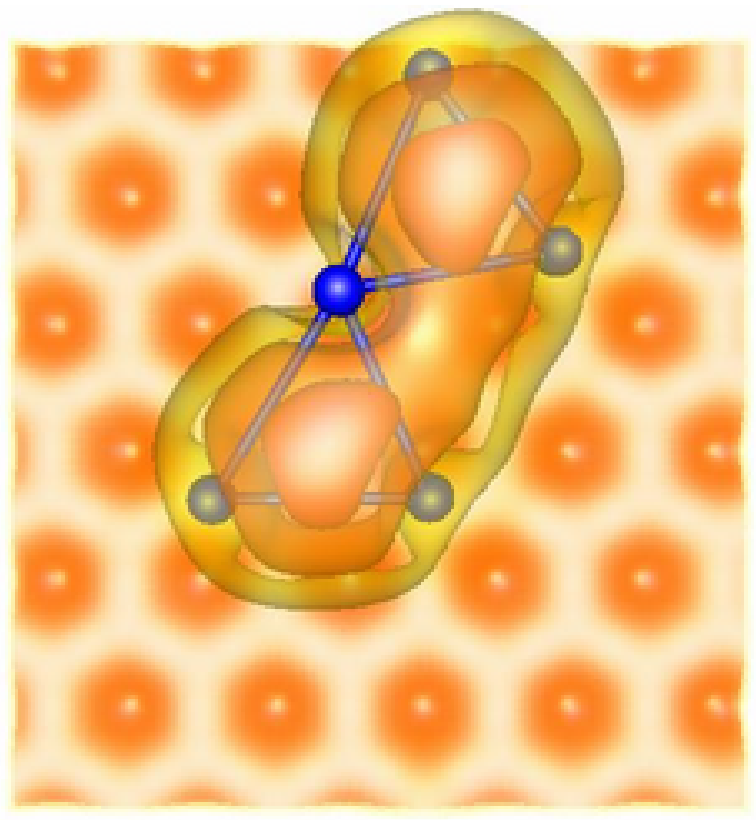}
\caption{Three density isosurfaces for Na$_5$-HOPG system. The corresponding
density values are 0.002 (yellow), 0.004 (orange), and 0.007 au (red),
respectively. The accumulated charges within the cluster are 2.37,
1.44, and 0.35 e, respectively.}
\label{vis_na5}
\end{figure}

The adsorption of Na$_5$ leads to a significant distortion from
planarity, with the central Na atom being much farther from the
surface (0.69 {\AA}) than the other atoms. Simultaneously, the
longest Na-Na bond is broken (4.11 {\AA}, see Table \ref{tab3}),
and the resulting C$_s$ structure (Fig. \ref{vis_na5}) comprises
two identical triangles connected via their apices. As for the
other clusters, changes in bond lengths are considerable, and
there are also changes in bond angles. The Na atoms are
coordinated with the surface in three ways: the central atom is on
a hollow site, the two atoms that initially comprised the broken
Na-Na bond sit on top of C-C bonds, and the two corner atoms are
directly above C atoms. Here, we have optimized the cluster
geometry with respect to a substrate consisting of two graphene
layers alone (120 C atoms). \cite{Gr2} The obvious changes are 
an increased separation of the middle Na atom from the surface (0.20 
{\AA}) and a further elongation of the longest (broken) Na-Na bond 
(0.11 {\AA}). The other bond distances and the formation energy 
$\Delta E$ are unchanged.

The electron density isosurface plot of Na$_5$-HOPG in Fig.
\ref{vis_na5} illustrates how the density is distributed within
the Na$_5$ cluster. The largest values are obtained inside the two
remaining triangles, but there is a component also in the
interstitial region next to the broken (or elongated) Na-Na bond.
The Na atom in the middle has a pronounced hole in the density,
but an atom-centered integration of charge density within a small
spherical volume ($R=1.5$ {\AA}) gives similar results (charges)
for each Na atom. This is explained by the fact that the most
coordinated Na atom has density contributions from both triangles.
Presumably, this atom prefers a larger distance from the surface
and a hollow site because of its higher Na coordination, whereas
the lower coordinated Na atoms tend to acquire positions closer to
carbons and C-C bonds. The same applies to Na$_4$ but on a smaller 
scale (see Fig. \ref{vis_na4}).

We have listed on Table \ref{tab4} formation energies for different
cluster/atom products on top of the HOPG substrate, calculated
assuming an initial state of $N$ free Na atoms in the gas phase. The
most obvious feature is the small formation energy of Na$_2$ containing 
products. This is caused by the full valence 
electron shell of Na$_2$ that reduces binding with the substrate. 
Even two separated Na atoms 
are more stable on graphite than a dimer. For larger systems Na$_3$, 
Na$_4$, and Na$_5$ are favored, indicating clustering processes, and 
the product Na$_3$+Na is slightly higher in formation energy than 
Na$_4$. The unpaired electron on the outermost 
shell of Na$_3$ (and Na) increases binding with HOPG as seen in Table
\ref{tab2}. A similar conclusion can be made about the high
stability of Na$_5$. This indicates that open shell clusters can
be more abundant on graphite than in the gas phase.

The weak binding of Na$_2$ compared to two separated atoms has
suggested to us to investigate the breaking of this bond. For
this purpose the Na$_2$ bond distance has been increased gradually
up to a point where the periodic Na(II) phase is obtained. Each
configuration has been optimized with respect to the surface, and
the total energy is calculated with the 5$\times$5$\times$1 $\bf
k$-point mesh. Our results show a monotonic increase
up to Na(II), which is the upper limit of Na-Na distance (6.51
{\AA}) in the supershell chosen. At this point, the energy is 0.30
eV higher, which should be considered as the lower bound of the
Na$_2$ dissociation energy on HOPG. This is still significantly
less than the gas phase value 0.68 eV, but the substrate now
causes the interaction between Na atoms to be long-ranged. On the
other hand, Na$_2$ stretches readily; as the Na-Na separation is
increased to 4.26 {\AA} where both atoms sit on a hollow site
(second nearest hexagons) the total energy change is only 0.09 eV,
but the distance $d_\bot$=3.09 {\AA} is 0.86 {\AA} less. This
suggests that Na$_2$ on graphite has very low frequency vibrational 
modes in both lateral and perpendicular directions.

\begin{figure}
\epsfig{file=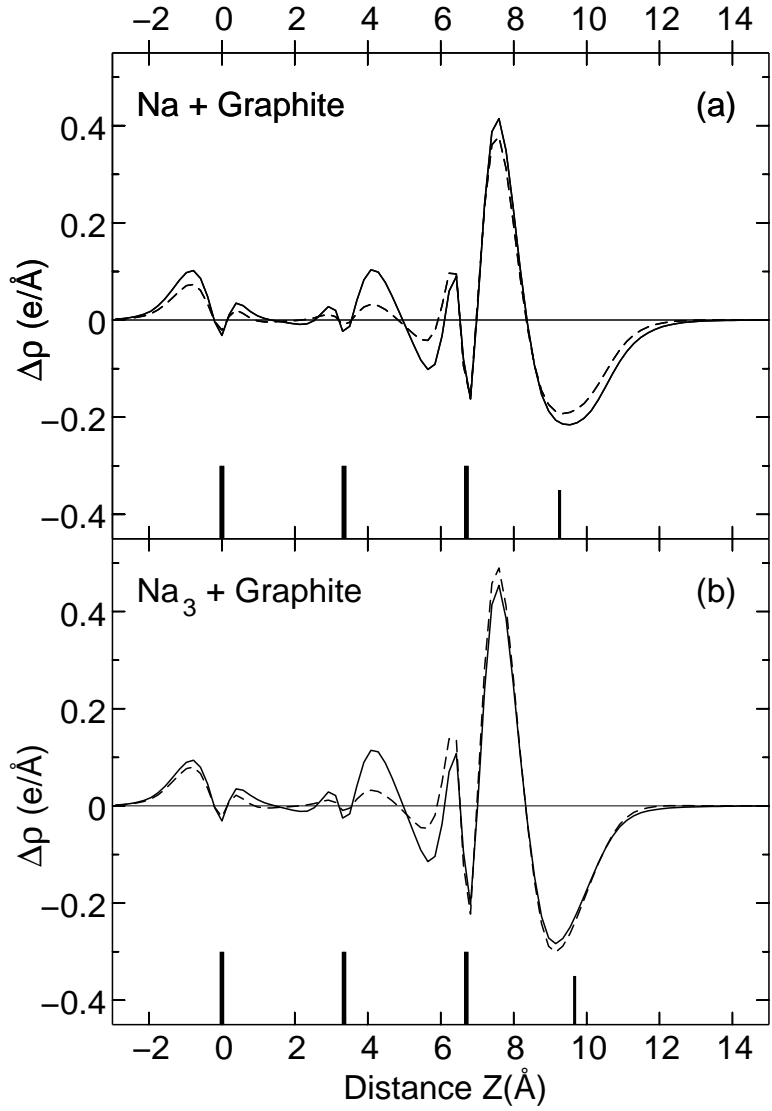}
\caption{Laterally averaged charge density difference for (a) Na-HOPG and (b)
Na$_3$-HOPG systems. The solid and dashed lines mark the 5$\times$5$\times$1
{\bf k}-point mesh and the $\Gamma$-point approximation, respectively. The
thick vertical bars denote the positions of graphene layers (longer bars)
and Na/Na$_3$ (shorter bar).}
\label{cdif}
\end{figure}

Charge transfer between the adsorbate and the substrate is studied
in detail in the case of Na-HOPG and Na$_3$-HOPG, \cite{Cha} and
the laterally averaged charge density differences ($\Delta\rho$)
are presented in Fig. \ref{cdif}. In both cases, the oscillating
profile of $\Delta\rho$ shows that the presence of adsorbate
affects the whole system including the lowermost (third) graphene
layer. The negative node close to the Na/Na$_3$ indicates a charge
transfer to the substrate that is partially counterbalanced by the
strong positive peak next to the first graphene layer (GR1, see
Table \ref{tab5}). The location and shape of the negative node is
different for Na and Na$_3$: for a single atom the charge is depleted
throughout the whole atomic volume causing a broad minimum in
$\Delta\rho$, whereas for Na$_3$ the minimum is deeper and biased
to the lower side of the cluster. Integration over this area gives
values $\Delta q=-0.47$ e and $\Delta q=-0.48$ e for Na and
Na$_3$, respectively. The similar $\Delta q$ values indicate that
the substrate does not support more excess charge, and it explains
the increased $d_\bot$ for Na$_3$, Na$_4$, Na$_5$ and (2$\times$2)
Na monolayer.

A layer-by-layer analysis of the graphite substrate in Table
\ref{tab5} shows that the charge transferred is distributed over
the three layers. In comparison with the middle layer (GR2),
$\Delta q$ is slightly larger for the lowermost layer (GR3). This
is probably a finite-size effect, a conclusion that is supported
by the $\Delta\rho$ profile. The inclusion of {\bf k}-points leads
to more pronounced oscillations near GR2 and GR3, whereas GR1 has
more accumulated charge in the $\Gamma$-point approximation. This
shows that the charge transferred becomes more delocalized as {\bf
k}-points are introduced in the lateral dimension. The $\Gamma$-point
approximation underestimates the amount of charge transfer also for 
Na, whereas for Na$_3$ the values are similar. Lamoen
and Persson \cite{Lam98} found $\Delta q=-0.40$ e for a
(4$\times$4) K monolayer, which agrees with our result
$\Delta q=-0.39$ for a corresponding density of Na using a single
$\Gamma$-point (Table \ref{tab5}).

\begin{figure}
\epsfig{file=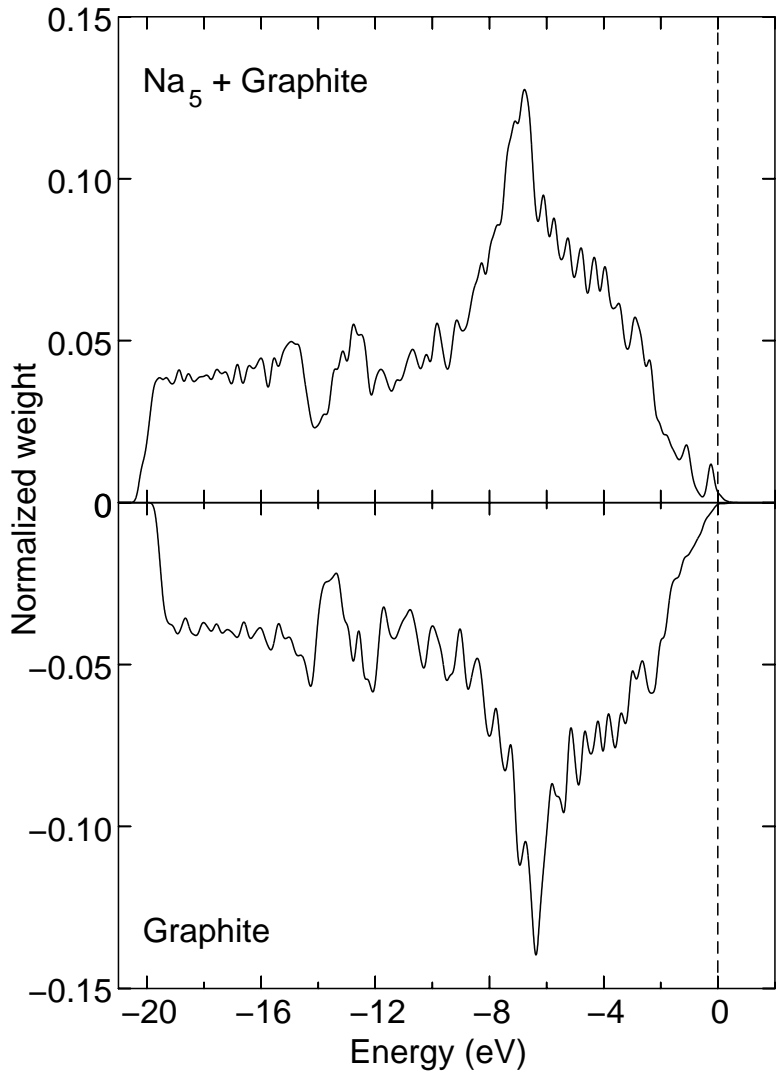}
\caption{Normalized DOS of Na$_5$-HOPG and HOPG systems calculated with
the 5$\times$5$\times$1 {\bf k}-point mesh. The data is interpolated to
correspond a 9$\times$9$\times$1 mesh, and Gaussians of 0.10 eV width were
used for each data point. The dashed line marks the Fermi level.}
\label{dos}
\end{figure}

The electronic densities of valence states (DOS) of Na$_5$-HOPG
and HOPG are plotted in Fig. \ref{dos}. The calculations were done
using a 5$\times$5$\times$1 Monkhorst-Pack $\bf k$-point mesh, and
the KS eigenvalues obtained are interpolated to correspond a
9$\times$9$\times$1 mesh (this resembles the common tetrahedron
method) \cite{Lam84}. The DOS of graphite substrate shows typical
features, \cite{Cha91,Cha92} including a steep rise at $-20$ eV due
to the 2D character of graphite, a dip at $-13$ eV after the first
two $\sigma$ bands, a large peak at $-6.5$ eV followed by a shoulder
in the decreasing profile with zero weight and zero gap at the
Fermi energy. Our substrate model then captures the relevant
properties of graphite, although the system is finite in the
perpendicular direction.

A very small effect is observed due to the Na$_5$ adsorption, and
the characteristic features of graphite substrate are clearly
visible. The conduction band is now being filled by the Na$_5$
valence electrons, which can be seen as a small peak at the Fermi
energy. The band structure of the three (spin-degenerate)
Na$_5$-HOPG conduction states reveals that the dispersion of the
first does not correspond to its graphite counterparts ($\pi^*$
bands), but resembles more the valence states ($\pi$ bands). This
is not true for the two other conduction states, where the lower
one shows only minor variation as a function of $\bf k$, and the
higher one resembles closely the graphite conduction bands. The two 
uppermost valence states are also affected by the presence of
Na$_5$, which can be seen as smaller dispersion. Together with the
lowest conduction state, this results in a small hump in the DOS
next to the minimum separating conduction and valence bands.

\section{Conclusion}

We have studied Na atoms and small Na clusters ($N\le5$) on
graphite using a DFT method that uses pseudopotentials and
periodic boundary conditions. In order to obtain reliable results
the simulated slab of graphite consists of three graphene layers,
and is sufficiently large to yield an appropriate separation
between the adsorbate replica in lateral dimension. In addition, a
high kinetic energy cutoff (70 Ry) for the plane wave basis set
and {\bf k}-points make the calculations very demanding in terms
of CPU time and memory.

For a dispersed phase, an Na atom has an adsorption energy of 0.51
eV at the hollow site 2.45 {\AA} above the surface. The small
diffusion barrier of 0.06 eV shows that the potential energy
surface of the Na atom is flat. These results are similar to the
recent results for NaC$_{60}$ compounds, where an adsorption
energy of 0.65 eV and a diffusion barrier of 0.07 eV were observed
at the hexagonal site. \cite{Roq03} A higher Na coverage leads to
a decreased interaction with the substrate as shown for the
(2$\times$2) monolayer ($d_\bot=3.21$ {\AA}). The dispersed phase
and (2$\times$2) monolayer differ little energetically, and
neither is found to be stable experimentally. Instead, the
calculated cluster formation energies favor clustering processes
(island formation) in agreement with experiment.
\cite{Whi94,Bre01,Bre03} The formation energies of the open shell
systems Na, Na$_3$, and Na$_5$ are larger than those of closed
shell cases Na$_2$ and Na$_4$. This is related to the
spin-degeneracy of the highest molecular orbital (odd-even
staggering) and, in contrast to free metal clusters, gives
rise to increased stability of odd cluster sizes on HOPG.

A charge density analysis for Na and Na$_3$ shows that
approximately 0.5 electrons are transferred to the substrate in
both cases, indicating that HOPG does not support much excess
charge, and that polarization effects weaken as the Na
coverage is increased. As shown before for K,
\cite{Anc93,Whi94,Lam98} this leads to decoupling between the
adsorbate and substrate, and a two-dimensional metallic film on
HOPG results. In the case of Na clusters, the partial loss in
electron density is evident in significant changes in cluster
geometries. For example, Na$_3$ is more like a closed shell
Na$_3^+$ ion, and consequently, the geometry is closer to an
equilateral triangle than that of a free Na$_3$. An interesting
observation is that the planarity of Na$_4$ and Na$_5$ is broken
as the atoms having more Na-Na bonds move farther from the
surface. Whether this is related to the experimentally observed
buckling of Na overlayers remains an open question.
\cite{Bre01,Bre02}

\section{Acknowledgments}

This work has been supported by the Academy of Finland under the
Finnish Centre of Excellence Programme 2000-2005 (Project No.
44875, Nuclear and Condensed Matter Programme at JYFL). The 
calculations were performed on the IBM-SP4 computers at the Center
for Scientific Computing, Espoo, Finland. J.A. has been supported
by the Bundesministerium f\"ur Bildung und Forschung (BMBF), Bonn,
within the Kompetenzzentrum Materialsimulation, 03N6015. We thank
R.O. Jones for valuable discussions and critical reading of the
manuscript.


\begin{table}
\caption{Na atom on graphite (0001) at different locations. $\Delta E_\bot$ is
calculated for both the 2$\times$2$\times$1 and 5$\times$5$\times$1 {\bf k}-point
meshes, but the geometries are optimized using the 2$\times$2$\times$1 mesh alone.
The vertical distance from the graphite layer ($d_\bot$), the Na-C distance, and
the carbon coordination number ($N_C$) are given also.}
\label{tab1}
\begin{center}
\vspace{2pt}
\begin{tabular}{c c c c c}  \hline

Point  & $\Delta E_\bot$ (eV)  & $d_\bot$ (\AA)  & Na-C (\AA) & $N_C$ \\  \hline
 0     & 0.12/0.51           & 2.45         & 2.83       &  6     \\
 1     & 0.08/0.47           & 2.49         & 2.59, 2.78 &  3     \\
 2     & 0.06/0.44           & 2.53         & 2.53       &  1     \\
 3     & 0.07/0.45           & 2.53         & 2.63       &  2     \\
 4     & 0.06/0.44           & 2.54         & 2.54       &  1     \\
 5     & 0.09/0.47           & 2.50         & 2.60, 2.78 &  3     \\
\hline

\end{tabular}
\end{center}
\end{table}

\begin{table}
\caption{Na atoms and clusters on graphite (0001). $\Delta E$ and $\Delta E_\bot$ are
calculated for both the 2$\times$2$\times$1 and 5$\times$5$\times$1 {\bf k}-point
meshes, but the geometries are optimized using the 2$\times$2$\times$1 mesh alone.}
\label{tab2}
\begin{center}
\vspace{2pt}
\begin{tabular}{c c c c c}  \hline
  & $\Delta E$/atom (eV) & $\Delta E_b$/atom & $\Delta E_\bot$/atom & $d_\bot$ (\AA) \\
\hline
Na(I)          &  0.12/0.51   &   ---      &  0.12/0.51 & 2.45        \\
Na(II)         &  0.15/0.33   &   0.09     &  0.07/0.24 & 2.81        \\
Na(III)        &  0.46/0.57   &   0.19     &  0.27/0.38 & 3.21        \\
Na$_2$         &  0.40/0.48   &   0.34     &  0.06/0.14 & 3.95        \\
Na$_3$         &  0.56/0.68   &   0.35     &  0.21/0.34 & 2.95, 2.98  \\
Na$_4^\dagger$ &  0.65/0.64   &   0.45     &  0.20/0.19 & 2.88, 3.12  \\
Na$_5^\dagger$ &  0.68/0.71   &   0.46     &  0.22/0.25 & 3.08, 3.77  \\
\hline
\end{tabular}
$^\dagger$larger substrate of 60 atoms per layer and simulation box of
12.30$\times$12.79$\times$16.70 \AA$^3$
\end{center}
\end{table}

\eject

\begin{table}
\caption{Optimized structures of adsorbed Na clusters. Distances in {\AA}ngstr\"om
and angles in degrees. The values in parentheses refer to the gas phase
structures.}
\label{tab3}
\begin{center}
\vspace{2pt}
\begin{tabular}{c c c c c}  \hline

         & Na$_2$        & Na$_3$        & Na$_4$        & Na$_5$      \\ \hline
Na-Na    & 3.07 (3.05  ) & 3.35 (3.17)   & 3.53 (3.43)   & 3.24 (3.33) \\
         &               & 3.26 (3.97)   & 3.27 (3.02)   & 3.34 (3.42) \\
         &               &               &               & 3.62 (3.36) \\
         &               &               &               & 4.11 (3.46) \\
\hline
$d_\bot$ & 3.95          & 2.95, 2.98    & 2.88, 3.12    & 3.08, 3.77  \\
\hline
Na-C     & 4.52          & 3.10          & 2.92, 3.30    & 3.09-4.12   \\
\hline
Angle    &               & 58.2 (77.7)   & 55.2 (52.2)   & 58.1 (61.8) \\
         &               &               & 62.4 (63.7)   & 66.7 (59.5) \\
         &               &               & 124.2 (127.8) & 55.3 (58.7) \\
         &               &               &               & 75.9 (60.5) \\
         &               &               &               & 155.2 (177.7) \\
\hline
Torsion  &               &               & 8.7           & 14.9, 30.7  \\
\hline
\end{tabular}
\end{center}
\end{table}

\begin{table}
\caption{Formation energies of Na products on graphite (0001). A large separation 
of end products is assumed.}
\label{tab4}
\begin{center}
\vspace{2pt}
\begin{tabular}{c c c}  \hline
Reactants (free)   & Products (graphite)  & $\Delta E$ (eV)  \\  \hline
Na                 & Na                   & 0.51 eV \\
\hline
2$\times$Na        & Na$_2$               & 0.96 eV \\
                   & 2$\times$Na          & 1.02 eV \\
\hline
3$\times$Na        & Na$_3$               & 2.05 eV \\
                   & Na$_2$ + Na          & 1.47 eV \\
                   & 3$\times$Na          & 1.53 eV \\
\hline
4$\times$Na        & Na$_4$               & 2.55 eV \\
                   & Na$_3$ + Na          & 2.56 eV \\
                   & Na$_2$ + Na$_2$      & 1.92 eV \\
                   & Na$_2$ + 2$\times$Na & 1.98 eV \\
                   & 4$\times$Na          & 2.04 eV \\
\hline
5$\times$Na        & Na$_5$               & 3.54 eV \\
                   & Na$_4$ + Na          & 3.06 eV \\
                   & Na$_3$ + Na$_2$      & 3.01 eV \\
                   & Na$_3$ + 2$\times$Na & 3.07 eV \\
                   & 5$\times$Na          & 2.55 eV \\
\hline
\end{tabular}
\end{center}
\end{table}

\begin{table}
\caption{Charge transfer in Na-HOPG and Na$_3$-HOPG (in electrons). The values
in parenthesis are for the $\Gamma$-point approximation.}
\label{tab5}
\begin{center}
\vspace{2pt}
\begin{tabular}{c c c c c}  \hline

            &  Na           &  GR1        &  GR2        &  GR3        \\  \hline
Na-HOPG     & -0.47 (-0.39) & 0.25 (0.28) & 0.09 (0.03) & 0.13 (0.09) \\
Na$_3$-HOPG & -0.48 (-0.49) & 0.26 (0.37) & 0.10 (0.03) & 0.12 (0.09) \\
\hline
\end{tabular}
\end{center}
\end{table}


\begin{thebibliography}{30}
\bibitem{Gr3}   Especially the distance from the surface is sensitive to the
number of graphene layers. For two graphene layers the distance is 0.16 {\AA}
(6.6 \%) larger than for three and four layers.
\bibitem{Cha}   The value of charge transfer can depend on the method
used for its determination. We use the same method as the previous
studies on alkali-HOPG systems. \cite{Anc93,Whi94,Lam98}
\end{thebibliography}
\end{document}